\documentclass[preprint, floats,amsmath,amssymb,preprintnumbers,superscriptaddress,floatfix]{revtex4-1}
\usepackage{graphicx}
\usepackage{bm}
\usepackage{color}
\usepackage{hyperref}
\usepackage{ulem}
\usepackage{amsmath}
\usepackage{amssymb}
\usepackage{mathtools}
\usepackage{mathrsfs}
\usepackage[T1]{fontenc}

\usepackage{here}
\usepackage[dvipsnames]{xcolor}

\usepackage{lipsum}
\usepackage{multirow}
\usepackage{array,booktabs}
\newcolumntype{M}[1]{>{\centering\arraybackslash}m{#1}}

\newcommand{\cmp}{{\bf CM$_p$} }
\newcommand{\cmn}{{\bf CM$_N$} }
\newcommand{\ie}{{\it{i.e.}}}

\newcommand{\mr}{\mathbf r}
\newcommand{\mF}{\mathbf F}

\newcommand{\bfrm}{\boldsymbol{\mathbf{r}}^N}
\begin{document}

\title{Structure and dynamics of responsive colloids with dynamical polydispersity}

\author{Upayan Baul}
\affiliation{Applied Theoretical Physics-Computational Physics, Physikalisches Institut, Albert-Ludwigs-Universit\"at Freiburg, D-79104 Freiburg, Germany}
\author{Joachim Dzubiella}
\affiliation{Applied Theoretical Physics-Computational Physics, Physikalisches Institut, Albert-Ludwigs-Universit\"at Freiburg, D-79104 Freiburg, Germany}
\affiliation{Cluster of Excellence livMatS @ FIT - Freiburg Center for Interactive Materials and Bioinspired Technologies, Albert-Ludwigs-Universit\"at Freiburg, D-79110 Freiburg, Germany}

\begin{abstract}
Dynamical polydispersity in single-particle properties, for example a fluctuating particle size, shape, charge density, etc., is intrinsic to responsive colloids (RCs), such as biomacromolecules or microgels, but is typically not resolved in coarse-grained mesoscale simulations.  Here, we present Brownian Dynamics (BD) simulations of suspensions of RCs modeling soft hydrogel colloids, for which the size  of the individual particles is an explicitly resolved (Gaussian) degree of freedom and dynamically responds to the local interacting environment.  We calculate the liquid structure, emergent size distributions, long-time diffusion, and property (size) relaxation kinetics for a wide range of densities and intrinsic property relaxation times in the canonical ensemble. Comparison to interesting reference cases, such as conventional polydisperse suspensions with a frozen parent distribution, or conventional monodisperse systems interacting with an effective pair potential for one fixed size, shows a significant spread in the structure and dynamics.  The differences, most apparent in the high density regimes, are due to many-body correlations and the dynamical coupling between property and translation in RC systems, not explicitly accounted for in the conventional treatments. In particular, the translational diffusion in the RC systems is surprisingly close to the free (single RC) diffusion, mainly due to a cancellation of crowding and size compression effects.  We show that an effective monodisperse pair potential can be constructed that describes the many-body correlations reasonably well by convoluting the RC pair potential with the density-dependent emergent size distributions and using a mean effective diffusion constant. 
\end{abstract}

\maketitle

\section{Introduction}

Recently, responsive colloids (RCs)~\cite{Motornov,adaptive,shape,shape2,capsules,Richtering, Schurtenberger} and related responsive materials~\cite{origami, actuator, Heuser, Andreas}  have drawn a lot of attention in the physical sciences.  Large responsiveness is a feature displayed, for example, by solvated polymers that have broad and often polymodal conformational distributions.  In particular, the class of {\it thermosensitive} polymers (including biomolecules) typically displays a critical solution temperature (CST) at which the polymers relatively sharply switch between two different physicochemical states. The switch can be induced by stimuli such as temperature, pH, or the (osmotic) pressure of a cosolute~\cite{Stuart2010, Kawasaki1997, Kawasaki2000, Sasaki1999}. The properties of a RC, made, for example, from a thermosensitive hydrogel, changes substantially at the CST: size changes by a factor of two or three are very typical~\cite{Richtering,Schurtenberger}.  The broad conformational and responsive distributions of polymers can be harvested for tailoring functionality in applications, for instance, actuators, soft sensors, triggered drug release by nano-carriers~\cite{Stuart2010,drug}, or selective catalysis in polymeric nanoreactor particles~\cite{Rafa1,msde2020modeling}. 
 
Typical examples of soft and functional RCs are thus hydrogel or block-copolymer particles synthesized with stimuli-responsive polymers~\cite{Stuart2010,Richtering,shape}, as well as biomolecular (or bio-inspired) polymeric particle assemblies from DNA, peptides, and proteins~\cite{protein,IDP_switching,DNA}. Many properties of such a RC, for example the macromolecular conformation~\cite{protein,RC,IDP_switching,chiwu}, size~\cite{matthias,denton:softmatter2016,Schurtenberger}, shape~\cite{denton:JCP2014,denton:JCP2016,  denton:softmatter2015,shape,shape2,Schurtenberger,hartmut_pnipam}, charge density~\cite{denton:softmatter2018,Schurtenberger,arturo}, etc., are thus responsive and highly fluctuating quantities. In other words, a single RC is intrinsically polydisperse with broad distributions, and any observable property $\sigma$ of a given RC follows a (parent) probability distribution, $p(\sigma)$.   A simple and typical example for $\sigma$ is the radius of gyration of a polymer~\cite{Fixman1962,kremer,matthias} or hydrogel~\cite{Richtering, Scotti, Winkler,ZaccMacromol2019,Schurtenberger}.   A functional response of a RC to the interacting environment means that the single-particle property distribution, $p(\sigma)$, and corresponding means and fluctuations will be modified and fed back to the spatial structure and dynamics of the whole suspension. Hence, if the property fluctuations are large and are related to function, it is clearly important to resolve the property dynamics explicitly in any modeling effort. 

In standard theoretical studies of the structure of suspensions of soft colloids the variable characterizing a property is not explicitly  resolved, i.e., all microscopic degrees of freedom are only implicitly contained in the coarse-grained effective pair potential between the colloids~\cite{likos, SoftColloid, nico1}.  Notable exceptions in the literature that included a property response are the works by Denton and Schmidt on the structure of colloid-polymer mixtures with compressible polymers~\cite{matthias} and more recent works by Denton {\it et al.} on penetrable and shape-fluctuating polymers and compressible hydrogels~\cite{denton:JCP2014, denton:JCP2016,denton:softmatter2015,denton:softmatter2016}. In these works, the polymeric size and/or shape was considered as a specific property degree-of-freedom and its distribution and bulk response to the surrounding cosolute was explicitly taken into account by an additional energetic one-body term in the Hamiltonian, either in a density functional theory (DFT) framework~\cite{matthias}  or in Monte-Carlo (MC) simulations~\cite{denton:JCP2014, denton:JCP2016,denton:softmatter2015,denton:softmatter2016}. From a different but related perspective, internal degrees of freedom have been recently discussed in so-called 'ultra-coarse graining' of small complex molecules with distinct internal states~\cite{ucg1,ucg2,ucg3}, for coarse-grained models of polymer beads~\cite{kremer} or protein domains~\cite{briels}, as well as a generic source for internal entropy driving demixing in coarse-grained systems~\cite{gobbo}. 

Recently, we presented a general statistical mechanics framework for the explicit resolution of properties of RCs and for studying the RC liquid structure and property response under the action of external fields or inhomogeneous environments~\cite{Lin:PRE}. We introduced an additional "property" degree of freedom as a collective variable in a formal coarse-graining procedure. The latter leads to an additional one-body term in the coarse-grained (CG) free energy, defining a single-particle "parent" property distribution $p(\sigma)$ for an individually polydisperse RC. We argued that in the equilibrium thermodynamic limit the structure of such a CG system of RCs is the same as that of a conventional polydisperse system of non-responsive particles. We discussed how an ideal gas of RCs distributes in external fields and inhomogeneous environments and also studied the effects of particle-particle interactions on one-body density profiles in approximate ways. However, to access one-body and pair structures as well as the coupled dynamics in these systems, more elaborate theories have to be developed or computer simulations have to be devised. In general, it is unclear how good dynamic properties are actually represented in coarse-grained simulations~\cite{nico2}. 

In this work, we use Brownian Dynamics (BD) computer simulations of RCs to study the bulk structure in terms of one- and two-body distribution functions and, in particular the (overdamped) dynamics of the translational and property degrees of freedom. For simplicity, we 'ad-hoc' suggest standard Brownian dynamics equations of motions in which the translation and property dynamics  can have individual relaxation time scales (that is, individual diffusion constants) and are coupled via the total RC Hamiltonian, similarly as suggested in previous work on coarse-grained protein simulations~\cite{briels}. As a specific case study, here we focus on a model of soft colloidal microgels and consider the colloidal size as the responsive variable, i.e., serving as an additional degree of freedom $\sigma$. It was shown recently that the size distribution of a single microgel can be conveniently expressed by a Gaussian distribution (reflecting essentially linear elasticity) and a Hertzian pair potential~\cite{SchurtNatComm2018,ZaccMacromol2019}. Such a system of RCs features a {\it dynamical polydispersity} in contrast to conventional polydisperse systems~\cite{Salacuse1,Salacuse2, Briano1984, Barrat1986, Kofke1988, bartlett, Poon1998, Warren, IgnacioPRL2000,  Sollich2002, schmid} in which a macromolecule's individual property is fixed for all times.  In particular, for the overdamped translational dynamics in the {\it canonical} simulation ensemble~\cite{BDpoly}, significant differences are expected due to the dynamic couplings between the fluctuating property and the translational degrees of freedom.  To elucidate clearly the novel structural and dynamical features of Brownian RCs, we compare the RC systems to known references cases, such as the conventional polydisperse system (which is actually the non-ergodic limit of RCs with frozen properties) and the conventional monodisperse systems~\cite{likos, SoftColloid,SchurtNatComm2018,ZaccMacromol2019} in which the property distributions are integrated out and only a coarse-grained pair potential remains with a single property (size) value as input. 

\section{Methods}

\subsection{Basic statistical mechanics of RCs}

\subsubsection{Hamiltonian}

In our recent work~\cite{Lin:PRE} we reported a statistical mechanics framework suited for liquids of responsive particles (RCs) through the introduction of an additional  internal,  \textquoteleft property\textquoteright~degree of freedom $\sigma$. The latter can be any responsive property of an RC while we limit the current study to the responsive size of a particle, in particular, we have a soft spherical microgel in mind. In the following, we briefly outline the key aspects pertinent to the current study. 

\noindent For a liquid of $N$ interacting RCs occupying a three-dimensional volume $V$ with number density $\rho=N/V$ in absence of any external fields, the coarse-grained Hamiltonian can be defined under the two-body approximation as~\cite{Lin:PRE}
\begin{equation}
H ( \bfrm, \sigma^N ) \simeq  F_0(N/V)+\sum_i^N \psi(\sigma_i) + \frac{1}{2} \sum_{i\neq j}^N \phi(\mr_i,\mr_j; \sigma_i,\sigma_j). 
\label{eqn:FreeEnergy}
\end{equation}
and describes the (free) energy of the microstate of the configuration \{$\mr_i,\sigma_i$\}, where the $\mr_i$ and $\sigma_i$ are positions and properties of particle $i=1..N$, respectively.  The first term is a one-body volume term independent of the configuration \{$\mr_i,\sigma_i$\} and accounts for the kinetic degrees of freedom of RCs. The second term is explicitly related to the property values realized in a system configuration, and carries the one-body contribution to the energetic changes associated with changes in the property of particle $i$. As such, $\psi(\sigma)$ represents a property potential energy function (or energy landscape), described by
\begin{equation}
 \beta \psi (\sigma)= -\ln\,p(\sigma),
 \label{eqn:PsiSigma}
\end{equation}
where $p(\sigma)$ is the property probability distribution function of a single RC,  and $\beta = 1/k_BT$ the inverse thermal energy. We call $p(\sigma)$  a \textit{parent} property distribution since it describes the intrinsic internal polydispersity of a particle, which will be in general modified by the interaction with other particles (or under the action of external fields).   The final term represents the two-body, effective pair potential, $\phi(\mr_i,\mr_j; \sigma_i,\sigma_j)$, which is explicitly dependent on both the positions and the property of the interacting RCs. 

\subsubsection{Liquid structure of RCs} 
The one-body density-property distribution can be defined as~\cite{Lin:PRE}
\begin{equation}
\rho(\mathbf r,\sigma) = \left\langle \sum_{i=1}^N \delta(\mathbf r - \mathbf r_i)\delta(\sigma-\sigma_i)\right\rangle,
\label{eqn:PropDen1Body}
\end{equation}
where the angular brackets signify ensemble averages. The ensemble-averaged one-body number density, $\rho(\mathbf r)$, and property distribution, $N(\sigma)$, distributions follow by integrations over properties or space, respectively, according to 
\begin{subequations}
\begin{align}
 \rho(\mr) &= \int_{\Sigma} {\rm d}\sigma \rho(\mathbf r, \sigma) \label{eqn:Den1Body} \\
 N(\sigma) &= \frac{1}{N} \int_V {\rm d}\mr \rho(\mathbf r, \sigma)\,\,.
 \label{eqn:Prop1Body}
 \end{align}
\end{subequations}
Here $\Sigma$ is the volume of the configurational space realizable by $\sigma$. The distribution $N(\sigma)$ in eq.~(\ref{eqn:Prop1Body}) is normalized to unity by dividing by the total number of particles $N$, and represents the \textit{emergent} property distribution accounting for interactions among RCs.

\noindent
Analogously, the two-body density-property distribution $ \rho^{(2)}$ can be defined as an average over products of $\delta$-functions~\cite{Lin:PRE}
and is related to the normalized pair distribution function $g(\mathbf r,\mathbf r'; \sigma, \sigma') \equiv g(r; \sigma, \sigma')$ through
$g(r; \sigma, \sigma') = \rho^{(2)}(\mathbf r,\mathbf r'; \sigma, \sigma')/\rho(\mr,\sigma)\rho(\mr',\sigma')$.
In the homogeneous situation, $\phi(\mr_i,\mr_j; \sigma_i,\sigma_j) \equiv \phi(r; \sigma_i,\sigma_j)$ depends on distance $r=|\mr_j - \mr_i|$ only, and not on individual particle coordinates. Then, the denominator can be expressed as $\rho(\mr,\sigma)\rho(\mr',\sigma') = \rho_0^2 p(\sigma) p(\sigma')$, where $\rho_0$ is the constant bulk density. Using the low-density limit relation 
\begin{equation}
 \lim_{\rho_0\rightarrow 0} g(r;\sigma,\sigma') = \exp[-\beta \phi(r;\sigma,\sigma')],
\end{equation}
it can be shown that the density profile $\rho_0g(r)$ around a test particle of same kind resulting from property-resolved interactions $\phi(r;\sigma,\sigma')$ in the limit $\rho_0\rightarrow 0$ can be accounted for by a more coarse-grained (CG), pair potential $v(r) = -k_BT \ln g(r)$ that satisfies the relation
\begin{equation}
\beta v(r) = - \ln \int_{\Sigma} {\rm d}{\sigma'}\int_{\Sigma} {\rm d}{\sigma} p(\sigma')p(\sigma)\exp[-\beta \phi(r;\sigma,\sigma')]. 
\label{eqn:vr}
\end{equation}
In fact, $v(r)$ is the conventional effective pair potential used in the modeling of purely monodisperse soft colloids~\cite{likos,SoftColloid,ZaccMacromol2019} and is naturally obtained by integrating out the property degrees of freedom in the low-density (pair) limit. It just depends on a single (mean) monodisperse size of the particles. In our study, we will compare RCs to this case and refer to it as the conventional monodisperse limit \cmp with respect to the parent distribution $p(\sigma)$.  Eq.~(\ref{eqn:vr}) is constructed such that the liquid structure for both treatments (conventional Hamiltonian based on $v(r)$ versus RC Hamiltonian, eq.~(1)) is the same in the low density limit. For larger densities, deviations will arise due to the many-body correlations which in general are different between the two treatments. 

\subsection{Brownian dynamics simulations of RCs}
\subsubsection{Basic simulation method}
Having recalled the basic statistical formalism in the previous section, in the following we provide the details of numerical simulations used in this work. 
In liquids of RCs, the particles are generically immersed in a solvent and are performing Brownian motion. Hence, we use Brownian dynamics (BD) equations of motion~\cite{BerendsenBook} for the dynamical evolution in the canonical (NVT) ensemble. For simplicity, we neglect inertial terms and assume the overdamped translational dynamics for the positions is propagated by the discretized form of BD, as  
\begin{equation}
 \mr_i(t+\Delta t) = \mr_i(t) + \frac{D_T}{k_BT} \mF_i(t) \Delta t + \bm{\xi},
 \label{eqn:BDtrans}
\end{equation}
where $\Delta t$ is the simulation timestep and $D_T$ is translational diffusion coefficient. $\mF_i = -\nabla_i H$ is the translational force experienced by $i$-th particle from pairwise interactions with all other particles, where $H$ is the potential energy following a Hamiltonian of form (1) (specified for our particular system further below). The term $\bm{\xi}$ represents stochastic forces stemming from the solvent fluctuations. For the evolution along a Cartesian dimension $l=x,y,z$, the corresponding component $\xi_l$ is drawn from a Gaussian distribution with mean $\langle \xi_l \rangle = 0$ and variance $\langle \xi_l^2 \rangle = 2 D_T \Delta t$ which is strictly $\delta$-correlated in time, that is, obeying the standard fluctuation-dissipation theorem (FDT) and Markovian behavior~\cite{AllenTildesley}. $D_T$ in eq.~(\ref{eqn:BDtrans}) is related to the translations friction coefficient $\zeta_T$ through $D_T = k_BT / \zeta_T$, following Stokes-Einstein relation $\zeta \propto \sigma$, if $\sigma$ is a particle size. 

For consistency, we also assume overdamped dynamics for the evolution of the property $\sigma$, specifically
\begin{equation}
  \sigma_i(t+\Delta t) = \sigma_i(t) + \frac{D_{\sigma}}{k_BT} F^{\sigma}_i(t) \Delta t + \xi_{\sigma},
 \label{eqn:BDsigma}
\end{equation}
where $D_{\sigma}$ formally represents the property diffusion coefficient (as a measure of the property fluctuation time scale). We assume also Gaussian (white) noise $\xi_{\sigma}$ that follows $\langle \xi_{\sigma} \rangle = 0$ and the standard (FDT) $\langle \xi_{\sigma}^2 \rangle = 2 D_{\sigma} \Delta t$, being also $\delta$-correlated in time. Analogous to the translational case we assume the Einstein-relation $D_\sigma = k_BT/\zeta_\sigma$ to hold, where $\zeta_\sigma$ is the property friction constant. Consistent with translational forces, the property force $F^{\sigma}_i$, originating from the interactions with the other colloids in the suspension, is defined as $F^{\sigma}_i = -{\partial}H/{\partial \sigma_i}$. Note that unlike $\mF_i$, $F^{\sigma}_i$ has contributions from the one-body term in Hamiltonian (1), which acts as a \textit{restoring} force when the instantaneous size $\sigma_i$ deviates from the mean size of an RC in isolation, $\sigma_0$, defined in the next section.

\subsubsection{The RC model for soft and responsive microgels}

As a simple and intuitively realizable property, in this work we associate $\sigma$ with size measures of RCs, such as the radius of gyration, or the hydrodynamic radius as measured in experiments.  In this study, for simplicity we restrict ourselves to relatively simple, swollen microgels far away from any critical condition, and consider a Gaussian parent distribution, as 
\begin{equation}
p(\sigma) = \frac{1}{\delta \sqrt{2\pi}} \exp \left[-\frac{1}{2} \left(\frac{\sigma - \sigma_0}{\delta}\right)^2 \right],
\label{eqn:p_sigma}
\end{equation}
with mean $\sigma_0 = 1.0$, setting our unit length scale, and standard deviation $\delta = 0.2$. Physically, the Gaussian distribution can be considered as an approximation for size fluctuations around an equilibrium state of a single RC. It can be directly related to the linear elastic response, e.g., of a microgel, and was shown to account well for its size fluctuations~\cite{ZaccMacromol2019}. The distribution $p(\sigma)$ according to eq.~(\ref{eqn:p_sigma}) is shown in Fig.~1. 

For the pairwise interaction potential $\phi(\mr_i,\mr_j; \sigma_i,\sigma_j)$ we consider the Hertzian potential
\begin{equation}
 \phi(r; \sigma_i,\sigma_j) =  \epsilon \left( 1 - \frac{r}{\sigma_{ij}} \right)^{5/2} \Theta \left( 1 - \frac{r}{\sigma_{ij}} \right),
 \label{eq:pp}
\end{equation}
where $\Theta$ is the Heaviside step function, and $\epsilon$ dictates the strength of interaction, $r=|\mr_j - \mr_i|$ is the interparticle distance, and $\sigma_{ij}=(\sigma_i+\sigma_j)/2$. The Hertzian potential reflects the soft repulsive interactions between elastic colloids, and is widely used to describe swollen states in microgel suspensions~\cite{ZaccMacromol2019,DentonSM2016,SchurtNatComm2018,SchurtJCP2014}.  For our RC system we always chose $\beta \epsilon=500$ which is in the typical range for modeling real experimental systems~\cite{SchurtNatComm2018}. The pair potential according to eq.~(\ref{eq:pp}) is shown in Fig.~1 for the specific choice $\sigma_i=\sigma_j=\sigma_0$. Here, we need to note that the Hertzian pair potential derived in~\cite{ZaccMacromol2019} is a conventional pair potential ($v(r)$ in our notation, cf.~eq(\ref{eqn:vr})) where the properties have been integrated out. Strictly speaking, however, we need a pair potential $\phi$, where the one-body property distributions to the free energy are subtracted according to Hamiltonian (1), see also ref.~\cite{Lin:PRE}.  However, we will see that Hertzian potential convoluted over Gaussians (as used here) again can be fitted well by a Hertzian pair potential which justifies the use of such a form also for $\phi$. Note also that the assumption was made in eq.~(\ref{eq:pp}) that the cross-interactions are simply additive, i.e., $\sigma_{ij}=(\sigma_i+\sigma_j)/2$ which is generally not true. In general, the cross-interaction in $ \phi(r; \sigma_i,\sigma_j)$ has to be determined by a consistent coarse-graining procedure~\cite{Lin:PRE}.


%
\noindent
Dropping the first kinetic term in eq.~(\ref{eqn:FreeEnergy}) and additive constants, we can now specify the final form of our Hamiltonian, $H$, for the BD simulation of a system of $N$ RCs modeling soft microgel colloids, as
\begin{eqnarray}
\label{eqn:Hamiltonian}
 H(\bfrm, \sigma^N) &=& \sum_i^N \frac{k_BT}{2 \delta^2} (\sigma_i - \sigma_0)^2 +   \\
  &\frac{1}{2}& \sum_{i\neq j}^N \epsilon \left( 1 - \frac{r}{\sigma_{ij}} \right)^{5/2} \Theta \left( 1 - \frac{r}{\sigma_{ij}} \right). \nonumber
\end{eqnarray}
\begin{figure}[htbp!]
 \centering
 \resizebox{!}{0.5\textwidth}{\input{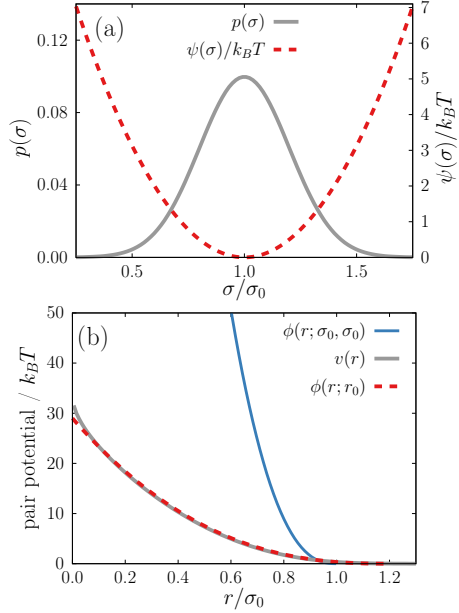}}
 \caption{{\footnotesize (a) Parent size distribution $p(\sigma)$ (gray, solid line) for a single RC and the corresponding one-body \textit{property} potential energy function $\beta\psi(\sigma) = - \ln p(\sigma)$ (red, dashed line), according to eq.~(\ref{eqn:p_sigma}). (b)
  The Hertzian pair potential for RCs, eq.~(\ref{eq:pp}), for the specific choice $\sigma_i=\sigma_j=\sigma_0$, $\phi(r; \sigma_0,\sigma_0)$ (solid blue line).     Also shown is the conventional monodisperse {\bf CM$_p$} pair potential, $v(r)$, using eq.~(\ref{eqn:vr}) (gray solid line) and its Hertzian pair potential approximation, $\phi(r;r_0)$ (red dashed line) used in the {\bf CM$_p$} Hamiltonian (eq.~\ref{eqn:CGvrHam}). 
 }}
 \label{fgr:figCGpot}
\end{figure}

\subsubsection{Mapping on conventional monodisperse (CM) pair potentials}

In eq.~(\ref{eqn:vr}) we showed that the conventional monodisperse ({\bf CM$_p$}) pair potential can be constructed by integrating out the property distributions $p(\sigma)$ in the low-density limit.  We observe that the numerically computed $v(r)$, being simply a double-convolution over Gaussians, is also well described by a Hertzian potential but with {\it rescaled monodisperse size} $r_0$, that is, $\phi(r;r_0,r_0)\equiv\phi(r;r_0)$. See Fig.~\ref{fgr:figCGpot} for a plot of the latter, which is much softer than the original Hertzian of the RCs for a fixed size, $\phi(r;\sigma_0,\sigma_0)$ from eq.~(\ref{eqn:Hamiltonian}).
This allows us to define the {\bf CM$_p$} Hamiltonian
\begin{eqnarray}
 H_{{\rm CM}_p}(\bfrm) &=& \frac{1}{2} \sum_{i\neq j}^N \phi(r;r_0) \nonumber \\
           &=& \frac{1}{2} \sum_{i\neq j}^N \epsilon_{CM_p} \left( 1 - \frac{r}{r_0} \right)^{5/2} \Theta \left( 1 - \frac{r}{r_0} \right)\;\;\;
 \label{eqn:CGvrHam}
\end{eqnarray}
with $\beta \epsilon_{CM_p} \approx 29.0$ and $r_0 \approx 1.20$. It is interesting to note that $r_0 \approx \sigma_0 + \delta$, i.e., the interaction length scale of the \cmp is the intrinsic mean size shifted by the width of the Gaussian in eq.~(\ref{eqn:p_sigma}) and with that 20\% larger. This will be a significant fact when we discuss effective Stokes radii for the self-diffusion of the RCs. 

We also phenomenologically attempt to consider many-body correlations in the \cmp model which become important for higher densities. Instead of convoluting over the low-density property distribution $p(\sigma)$, it feels reasonable to convolute over the emergent distributions $N(\sigma;\rho)$ (which are output from the simulations at density $\rho$).  Following the same methodology eq.~(\ref{eqn:vr}) as for $v(r)$, we can thus define density-dependent \cmn  pair potentials $\tilde{v}(r;\rho)$ for a conventional monodisperse system with respect to $N(\sigma)$, via
\begin{equation}
\beta \tilde{v}(r;\rho) = -\ln \int_{\Omega} {\rm d}{\sigma'}\int_{\Omega} {\rm d}{\sigma} N(\sigma';\rho)N(\sigma;\rho) e^{-\beta \phi(r;\sigma,\sigma')},
\label{eqn:vtilder}
\end{equation}
where we have replaced the single-particle property distribution $p(\sigma)$ in eq.~(\ref{eqn:vr}) by the emergent distribution $N(\sigma;\rho)$ at a given number density $\rho$. In the low-density limit (LDL),  $\rho \rightarrow 0$, thus $\tilde{v}(r;\rho) \simeq v(r)$. We mention in advance that the pair potential $\tilde{v}(r;\rho)$ is \textit{a priori} unknown, and we invoke it to test the potential of the coarse graining methodology at densities far from LDL, assuming that the emergent distribution is known (say from experiments, some theory, or directly from the RC simulations as in our case).  We observe that as with $v(r)$, $\tilde{v}(r;\rho)$ can also be described by  Hertzian potentials $\phi(r;r_0(\rho))$, whence a density-dependent energy function for a conventional monodisperse system can then be defined as the {\bf CM$_N$} Hamiltonian 
\begin{eqnarray}
&H_{{\rm CM}_N}&(\bfrm;\rho) = \frac{1}{2} \sum_{i\neq j}^N \phi(r;r_0(\rho)) \nonumber \\
           &=& \frac{1}{2} \sum_{i\neq j}^N U(\rho) \left( 1 - \frac{r}{r_0(\rho)} \right)^{5/2} \Theta \left( 1 - \frac{r}{r_0(\rho)} \right).
 \label{eqn:CGvetarHam}
\end{eqnarray}
The energy prefactor $U(\rho)$ is now dependent on density because the emergent distributions are density-dependent. We will determine the prefactor for every $\rho$ individually and summarize later in Tab.~II. 

\subsubsection{Tuning the dynamical coupling and reference cases}
\label{sec:dyn}
In our simulations, we set the translational diffusion coefficient for RCs with size $\sigma_0$ to unity, \ie, $D_T^0 = 1.0$, that is, it defines  the overall timescale unit in our simulations. To study systematically the effects of property dynamics on the translational diffusivity in the systems of RCs, we perform for each system two independent simulations. First, with a constant (size-independent) translational diffusion $D_T = D_T^0$, and also one with a (more realistic) $\sigma$-dependent diffusion, according to Stokes scaling, $D_T = D_T(\sigma) = \sigma_0 D_T^0/\sigma(t)$.  The property diffusion coefficient $D_{\sigma}$, setting the timescale of the property fluctuations, we always define relative to the unit $D_T^0$, as $D_{\sigma} = \alpha D_T^0$, with a dynamical scaling parameter $\alpha \geq 0$. Expressed by the friction constants, $\alpha = \zeta_T/\zeta_\sigma$. 

It is at this point important to note that the liquid structure, e.g., pair distributions, of an RC system, which are true equilibrium ensemble averages, should not depend on the dynamical variable $\alpha$. In other words,  for a system that samples full equilibrium (i.e., the simulation trajectory is ergodic), time and ensemble averages  are the same and should not depend on the time scales of the fluctuations of the degrees of freedom. 
However, dynamic quantities, on the other hand, such as the translational long-time diffusion, are expected to depend on $\alpha$ in the RC systems because of the dynamical couplings between translational and property motion. Also, extreme limits of $\alpha$, such as $\alpha\rightarrow 0$ could also be interesting to consider because ergodicity (i.e., full phase space sampling) is violated on the simulation time scale in the (finite) canonical ensemble and we recover other known reference limits. 
Hence, it is interesting to play with $\alpha$ and also to consider two limiting cases more explicitly: 

\begin{itemize}
\item
For $\alpha \rightarrow 0$, it follows $D_\sigma \rightarrow 0$, and the properties will never relax on the simulation time scale. Hence, we recover a conventional polydisperse system in the canonical ensemble: The properties are initially distributed according to a distribution $p(\sigma)$, but since the property degree of freedom is essentially frozen, the property distribution will never change in the simulation time window. Hence, the structure and dynamics in the simulation 'equilibrium' in the limit $\alpha=0$ is that of a conventional polydisperse ({\bf CP}) system in the canonical ensemble~\cite{BDpoly}.

\item For, $\alpha \rightarrow \infty$,  it follows $D_\sigma \rightarrow \infty$, and we consider a system with 'instantaneous' property response, that is the property fluctuations are quickly integrated out. Hence, it could be expected that, e.g.,  translational dynamic observables in equilibrium should be also well representable by a conventional monodisperse ({\bf CM}) system without internal property degree of freedom. Here, however, the questions arise what would be the effective pair potential and the effective translational diffusion constant, representing the dynamics of such a reference CM system well. As we will see, these questions arise actually for all values of $\alpha$. 

\end{itemize}
\begin{table*}[ht!]
  \begin{center}
  \begin{tabular}{@{}|c|M{30mm}|M{25mm}|M{30mm}|M{20mm}|M{20mm}|c|@{}}
    \hline \hline
    system & description & Hamiltonian & $\rho\sigma_0^3$ & $\alpha$ & $D_T$ & \#systems \\
    \hline \hline
    {\bf RC} & {\footnotesize liquids of RCs} & $H(\bfrm, \sigma^N)$, eq.~(\ref{eqn:Hamiltonian}) & {\footnotesize 0.019, 0.19, 0.57, 0.95, 1.33 } & {\footnotesize $10^{-3}$, $10^{-2}$, $10^{-1}$, 1, 10} & $D_T^0$, $D_T(\sigma)$ & 50 \\
    \hline
    {\bf CP} & {\footnotesize conventional polydisperse} & -- " --  &  -- " --  & 0 & -- " --  & 10 \\
    \hline
    \cmp & {\footnotesize conventional monodisperse using $p(\sigma)$} & $H_{{\rm CM}_p}(\bfrm)$, eq.~(\ref{eqn:CGvrHam}) & -- " --  & N/A & $D_T^0$, $D_T(r_0)$ & 10 \\
    \hline
    \cmn & {\footnotesize conventional monodisperse using $N(\sigma;\rho)$} & $H_{{\rm CM}_N}(\bfrm;\rho)$, eq.~(\ref{eqn:CGvetarHam}) & -- " --  & N/A & $D_T^0$, $D_T(r_0(\rho))$, $D_T(\bar{\sigma}(\rho))$, $D_T(\tilde{\sigma}(\rho))$ & 20 \\
    \hline \hline
  \end{tabular}
  \end{center}
  \caption{{\footnotesize Summary of the simulated systems, their Hamiltonians, and corresponding syntax: {\bf RC} are the responsive colloids which are focus of this study. {\bf CP} are conventional polydisperse colloids with frozen property dynamics. {\bf CM}$_p$ and {\bf CM}$_p$ are conventional monodisperse colloids where the property degrees of freedom have been integrated out using the parent or emergent distributions $p(\sigma$) or $N(\sigma)$, respectively.   $\rho\sigma_0^3$ is the dimensionless number density, and $\alpha = D_\sigma/D_T^0$} the ratio between the property and translation fluctuation time scales. The translational diffusion in the models, $D_T$, is taken either constant or size-dependent (Stokes), see text in sections~\ref{sec:dyn} and \ref{sec:sim}.}
  \label{tbl:Nomenclature}
\end{table*}
%

%

%
%
%
%

\subsection{Simulation details, parameters, and analysis}
\label{sec:sim}
As the above discussion describes, simulated systems in our study are characterized by (i) the Hamiltonian, (ii) the number density of particles $\rho\sigma_0^3$, (iii) the translational diffusion coefficient $D_T$, and for the RC systems (iv) the property diffusion coefficient scaling $\alpha$. We perform simulations at varying number densities, $\rho=N/V$, spanning the low density limit, $\rho\sigma_0^3 = 0.019$, to close packing, $\rho\sigma_0^3 = 1.33$, over a range of $\alpha$ values, namely $\alpha = 0, 10^{-3}, 10^{-2}, 10^{-1}, 1$, and $10$. As such, in terms of interplay between translational and size dynamics, we cover conventionally polydispersity ({\bf CP}) as a non-ergodic limit for $\alpha=0$, up to very fast dynamics in size compared to translation, for $\alpha=10$. 

For insightful comparisons, we also simulate the monodisperse systems \cmp and \cmn defined by eqs.~(\ref{eqn:CGvrHam}) and~(\ref{eqn:CGvetarHam}), respectively, for all values of $\rho\sigma_0^3$, using friction coefficients $D_T = D_T^0$, and also $D_T = D_T(x) = \sigma_0 D_T^0 / x$, where $x$ is given by the effective emergent size in the pair potentials, $r_0$ or $r_0(\rho)$, as applicable. For \cmn systems, two additional $D_T(x)$ with $x=$ ($\bar{\sigma}(\rho)$, $\tilde{\sigma}(\rho)$) are considered, given by 
\begin{equation}
 \bar{\sigma}(\rho) = \int_{\Omega} \sigma N(\sigma;\rho) {\rm d}{\sigma}
 \label{eqn:SigmaBar}
\end{equation}
and 
\begin{equation}
 \tilde{\sigma}(\rho)^{-1} = \int_{\Omega} \frac{1}{\sigma} N(\sigma;\rho) {\rm d}{\sigma}
 \label{eqn:SigmaTilde}
\end{equation}
respectively. Physically, they map to effective single particle sizes that result from averaging friction ($\propto \sigma$) and diffusion ($\propto \sigma^{-1}$) in the polydisperse RC systems. We will compare both routes and discuss their performance when compared to the RC system. 
In Tab.~\ref{tbl:Nomenclature} we provide a summary of the various systems simulated, and introduce nomenclature that will be used in the rest of the manuscript. 

\noindent
BD simulations were performed for all all 90 systems described in Tab.~\ref{tbl:Nomenclature} using eqs.~(\ref{eqn:BDtrans}) and~(\ref{eqn:BDsigma}) (when applicable). All systems comprised of $N=512$ particles, and periodic boundary conditions were used along all Cartesian directions. All simulations of polydisperse systems, with energy function defined by eq.~(\ref{eqn:Hamiltonian}), were initialized to have the parent $p(\sigma)$ distribution in particle sizes. Using a timestep $\Delta t = 10^{-4} \tau_{\mathrm{BD}}$, all systems were equilibrated for $100\tau_{\mathrm{BD}}$, followed by production simulations for $1000\tau_{\mathrm{BD}}$. Analysis were performed on the production simulation data, which was written every 1000th timestep. For a subset of $\mathbf{RC}$ systems, smaller ($250\tau_{\mathrm{BD}}$) production simulations were additionalally performed during which data were written every 10th timestep. The latter were used for analysis of property relaxation time-scales. Gaussian distributed random numbers were generated using the Marsaglia polar method using algorithm due to Knuth~\cite{KnuthBook,Marsaglia}. All simulations were performed under constant volume and temperature ($=1.0$), using self-written codes. 

\begin{figure*}[htb!]
 \centering
 \resizebox{!}{0.255\textwidth}{\input{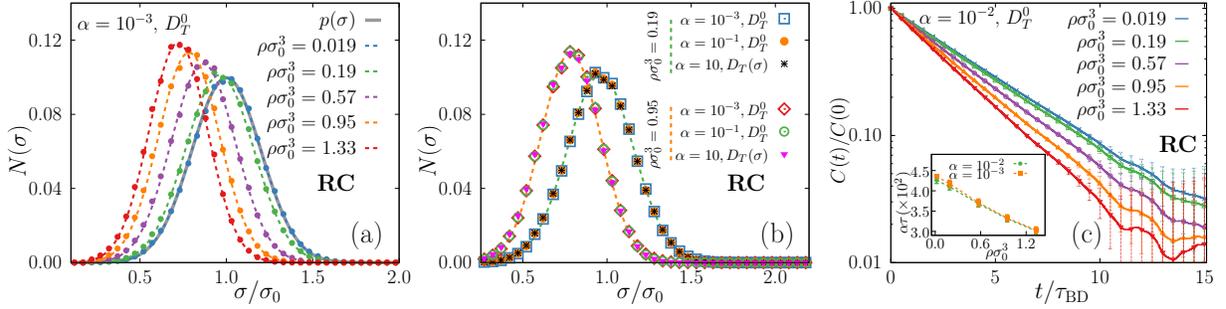}}
 \caption{{\footnotesize (a) Emergent size distributions, $N(\sigma)$,  for RCs with different bulk number densites $\rho\sigma^3$. The distributions fit to Gaussian functions for all $\rho$ which are shown with dashed lines. The single-particle distribution $p(\sigma)$ is shown with gray solid line. At the low-density limit (LDL) $\rho\sigma_0^3=0.019$, $N(\sigma) \simeq p(\sigma)$. (b) $N(\sigma)$ for RCs is independent of the dynamical parameters $\alpha$ and $D_T$. Four orders of magnitude in $\alpha$ are shown, together with both constant friction $D_T^0$ and instantaneous size dependent friction $D_T(\sigma(t;\sigma))$. For a given $\rho\sigma_0^3$, $N(\sigma)$ distributions can be observed to be in perfect agreement. Gaussian distributions from (a) are shown in (b) with dashed lines for $\rho\sigma_0^3=0.19$ (green) and $\rho\sigma_0^3=0.95$ (orange). (c) Log-lin plot showing the exponential decay of auto-correlation of RC size $C(t)$ defined by eq.~(\ref{eqn:CorrSigma}) for $\alpha = 10^{-2}$ for all $\rho$. The inset plot shows the decay of the relaxation time-scales ($\tau$) following $C(t) = C_0 \exp(-t/\tau)$ as a function of number density $\rho\sigma_0^3$. Curves for different $\alpha$ values collapse upon scaling $\alpha \tau$.}}
 \label{fgr:Nsigma_eta}
\end{figure*}

Regarding analysis, the radial pair distribution functions (RDFs), $g(r)$,  and the emergent size distributions, $N(\sigma)$, were computed as time averages over the equilibrium trajectories using standard histogramming methods.  Translational diffusion constants were evaluated using the mean squared displacement ($\mathrm{MSD}$), via 
\begin{equation}
 \mathrm{MSD}(t) = \frac{1}{N} \left\langle \sum_i \left( \mr_i(t_0+t) - \mr_i(t_0) \right)^2 \right \rangle,
 \label{eqn:MSDComput}
\end{equation}
where $\mr_i(t_0)$ and $\mr_i(t_0+t)$ are particle positions at times $t_0$ and $t_0+t$, respectively, and the average taken over all initial times generated from the snapshots of the production trajectory. The $\mathrm{MSD}$ is related to \textit{effective} translational diffusion coefficient $D_T^{\mathrm{eff}}$ in 3D through the $\mathrm{MSD}(t) = 6 D_T^{\mathrm{eff}}$. From our simulated data, we calculated $D_T^{\mathrm{eff}}$ using the long-time regime fit of $ \log_{10} \mathrm{MSD}(t) = a + \log_{10} t$, whence $D_T^{\mathrm{eff}} = 10^a / 6$. The error was estimated using $\Delta D_T^{\mathrm{eff}} = \left( 10^a \Delta a \ln 10 \right) / 6$, where $\Delta a$ is obtained from fit.

Since we consider RC systems with responsive, density-dependent sizes ,we need to define emergent packing fractions.  The average emergent packing fraction, $\eta$, is calculated using 
\begin{equation}
 \eta = \frac{1}{V}  \sum_i \frac{\pi}{6} \langle \sigma_i^3 \rangle = \rho\frac{\pi}{6}\sigma_{\rm eff}^3, 
 \label{eqn:ChiComput}
\end{equation}
with which a global, effective packing size $\sigma_{\rm eff}(\rho)$ can be defined. For the input packing fraction, $\eta_0$, used for a consistent initialization of systems, we used $\sigma_i=\sigma_0$ in the above equation. 

The property relaxation dynamics in $\mathbf{RC}$ systems were quantified using the auto-correlation function (ACF) $C(t)$, defined as
\begin{equation}
 C(t) = \langle \sigma(t) - \sigma_0' \rangle \langle \sigma(0) - \sigma_0' \rangle,  
 \label{eqn:CorrSigma}
\end{equation}
where $\sigma'_0$ is the mean of the emergent property distributions. A mono-exponential relaxation time ($\tau$) was estimated by fit to the functional form $C(t) = C_0 \exp(-t/\tau)$.

\section{Results}
\subsection{Emergent property distributions and relaxation}

\begin{table}[!htb]
  \begin{center}
  \begin{tabular}{@{}ccccccc@{}}
    \hline
    $\rho\sigma_0^3$ & $\eta_0$ & $\eta$ & $\sigma_0'/\sigma_0$ & $\delta'/\sigma_0$ & $r_0(\rho)$ & $U(\rho)$ \\
    \hline
    0.019 & 0.01 & 0.01 & 1.0 & 0.20 & 1.20 & 29.0\\
    0.19  & 0.10 & 0.09 & 0.961 & 0.197 & 1.15 & 27.0\\
    0.57  & 0.30 & 0.19 & 0.866 & 0.186 & 1.05 & 25.0\\
    0.95  & 0.50 & 0.24 & 0.784 & 0.176 & 0.96 & 23.0\\
    1.33  & 0.70 & 0.26 & 0.722 & 0.168 & 0.89 & 21.5\\
    \hline
  \end{tabular}
  \end{center}
  \caption{{\footnotesize Densities and descriptors for emergent size distributions of the $\mathbf{RC}$ systems. Symbols $\eta_0$ and $\eta$ represent input and emergent packing fractions, respectively. Note that $\eta$ for $\mathbf{CP}$ systems is the same as $\eta_0$. Parameters $\sigma_0'$ and $\delta'$ represent mean and variance of the $N(\sigma)$ distributions, cf. eq.~(\ref{eqn:N_sigma}). The fitted results hold for all simulated $\alpha>0$ and $D_T$. The $r_0(\rho)$ for the $\mathbf{CM}_N$ systems are calculated using the $N(\sigma)$ distributions and used in eq.~(\ref{eqn:CGvetarHam}). $U(\rho)$ values used in eq.~(\ref{eqn:CGvetarHam}) are also listed for completeness.}}
  \label{tbl:N_sigma}
\end{table}

\noindent 
We first discuss how density affects the property distributions and dynamics. Representative results for the emergent distributions, $N(\sigma)$, are shown in Fig.~\ref{fgr:Nsigma_eta}(a) for various $\rho\sigma_0^3$ with slowest simulated dynamics in size, $\alpha = 10^{-3}$. As can clearly be seen, with increasing density $\rho$ the $N(\sigma)$ distributions shift to smaller values of $\sigma$. In other words, the RCs respond to increased crowding by contracting in size. (This will not happen in the canonical {\bf CP} system where $p(\sigma)$ remains frozen.) This phenomenon is well known in the literature for microgels, where the \textit{swelling ratio} of microgels has been observed to decrease with density, see, e.g., reference~\cite{DentonSM2016}. Note that with $\sigma_0=1.0$, $\sigma / \sigma_0$ is indeed the swelling ratio, which is defined as the dimension of an RC relative to its dimension in isolation. We observe that the $N(\sigma)$ distributions at all $\rho$ can adequately be described by (see Fig.~\ref{fgr:Nsigma_eta}) Gaussian distributions
\begin{equation}
N(\sigma) = \frac{1}{\delta' \sqrt{2\pi}} \exp \left[-\frac{1}{2} \left(\frac{\sigma - \sigma_0'}{\delta'}\right)^2 \right].
\label{eqn:N_sigma}
\end{equation}

\begin{figure*}[htbp!]
 \centering
 \resizebox{!}{0.255\textwidth}{\input{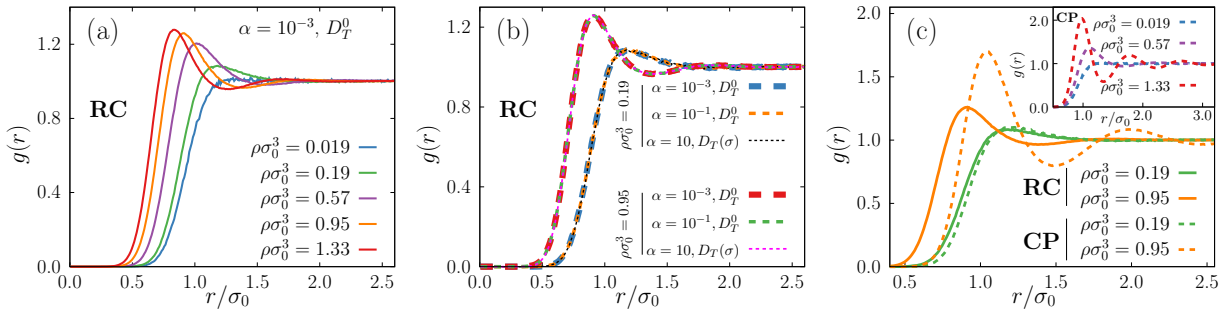}}
 \caption{{\footnotesize Radial pair distribution function $g(r)$ for $\mathbf{RC}$ systems as a function of (a) number density $\rho\sigma_0^3$ and (b) dynamical parameters $\alpha$ and $D_T$ at two values of $\rho\sigma_0^3$. As with $N(\sigma)$ in FIG.~\ref{fgr:Nsigma_eta}(b), $g(r)$ for RCs is also independent of dynamical parameters. (c) A comparison of $g(r)$ of $\mathbf{CP}$ (solid) and $\mathbf{RC}$ (dashed) systems at $\rho\sigma_0^3=0.19$ (green) and $0.95$ (orange). Inset plot of (c) shows $g(r)$ of $\mathbf{CP}$ systems at other simulated number densities.}}
 \label{fgr:gr_RC_CP}
\end{figure*}

The mean $\sigma_0'$ and standard deviation $\delta'$ for all densities $\rho$ are given in Tab.~\ref{tbl:N_sigma}, together with input ($\eta_{0}$) and emergent packing fractions ($\eta$). The latter substantially deviate from the input $\eta_0$ at high densites because of the large compression.  From principles of statistical mechanics, for $\alpha \neq 0$, the structure should be independent of $\alpha$, as well as $D_T$. Indeed,  Fig.~\ref{fgr:Nsigma_eta}(b) shows that $N(\sigma)$ at a given $\rho\sigma_0^3$ is independent of $\alpha$ and $D_T$.

In Fig.~2(c) we present the property ACFs which show a near-exponential decay for all densities. Only for longer times, $t \simeq 11-14\tau_B$ noticeable correlations (peak structures) occur which seemingly systematically grow with density. The source of these interesting long-time correlations are currently unclear and shall be studied more intensely in future work. From the exponential decay of the ACFs we can extract relaxation time scales $\tau$ as summarized in the inset to Fig.~2(c). The property relaxation time decreases (accelerates) with increasing density. This can be understood by the fact that the emergent, Gaussian-like distributions become narrower with increasing packing fraction, cf. $\delta'$ in Tab.~II. 

\subsection{Pair structure}

Let us now turn to the pair structure of the RC system, as exemplified in Fig.~\ref{fgr:gr_RC_CP} for various densities, $\rho$, and dynamical scaling, $\alpha$. 
As expected for increasing density and the compressed size distributions, the structural correlations grow and the peak position of the RDF, $g(r)$ shift to the left, see Fig.~\ref{fgr:gr_RC_CP}(a). Also as expected, Fig.~\ref{fgr:gr_RC_CP}(b) shows that the $g(r)$ is independent of $\alpha$ and $D_T$, i.e., all our RC systems are consistently simulated in full equilibrium and the property dynamical timescale plays no role. It is insightful to compare to the {\bf CP} system with frozen initial parent distribution of sizes in Fig.~\ref{fgr:gr_RC_CP}(c). Also for the {\bf CP}, the $g(r)$ changes with $\rho\sigma_0^3$, where larger density also implies greater structuring. However, the direct comparison to RCs in Fig.~\ref{fgr:gr_RC_CP}(c) demonstrates that only for low densities ($\rho\sigma_0^3 = 0.19$) the RDFs agree well, while for larger densites (here, $\rho\sigma_0^3 = 0.95$) the structure substantially deviates. The reason is clearly the frozen polydispersity in the canonical ($NVT$) ensemble which does not allow the {\bf CP} system to compress and develop an emergent property distribution.  

\noindent
\begin{figure*}[htbp!]
 \centering
 \resizebox{!}{0.5\textwidth}{\input{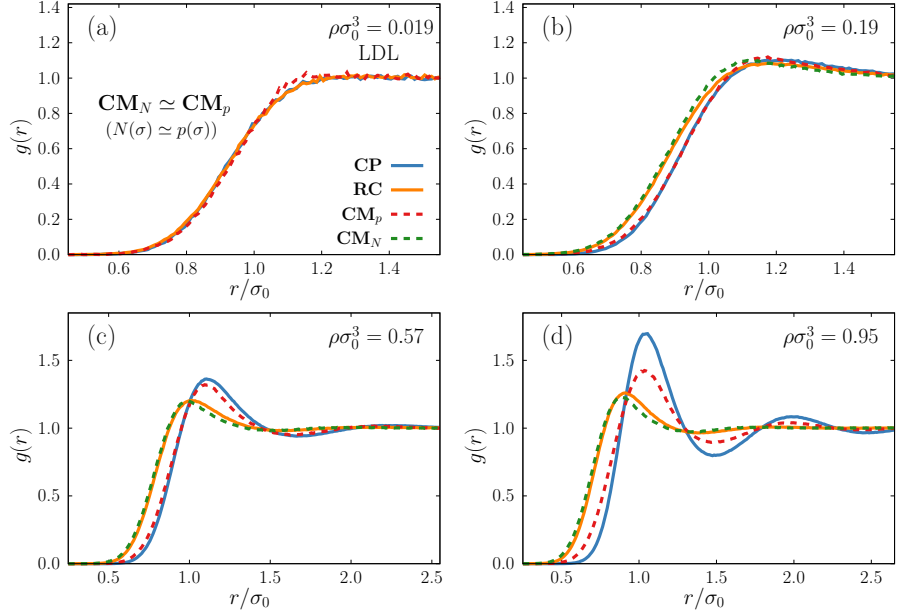}}
 \caption{{\footnotesize Comparison of $g(r)$ for the $\mathbf{RC}$ systems (orange solid line) with polydisperse systems  $\mathbf{CP}$ (blue solid lines) as well as coarse grained monodisperse systems (dashed lines) $\mathbf{CM}_p$ (red) and $\mathbf{CM}_N$ (green). (a) At LDL $\rho\sigma_0^3=0.019$, $\mathbf{CM}_p \simeq \mathbf{CM}_N$ since $N(\sigma) \approx p(\sigma)$), and the pair structures in all systems agree. (b)-(d): As $\rho\sigma_0^3$ increases, $\mathbf{CP}$ and $\mathbf{CM}_p$  more and more deviate from the structure of $\mathbf{RC}$, while $\mathbf{CM}_N$ can still describe the RC structure well for moderate densities, $\rho\sigma_0^3=0.57$.}}
 \label{fgr:CGstr}
\end{figure*}

We now compare the RCs to the conventional polydisperse $\mathbf{CP}$ and the monodisperse systems, \cmp and \cmn in Fig.~\ref{fgr:CGstr}.   First, in Fig.~\ref{fgr:CGstr}(a) we compare the structures at a density well representing the low-density-limit (LDL), $\rho\sigma_0^3=0.019$. In this limit, no many-body correlations occur and the pair structures in all systems agree and are given by the LDL $g(r)$.  However, away from the LDL, the emergent property distributions deviates from the parent, $N(\sigma) \neq p(\sigma)$, and structural differences should appear.  Indeed, Figs.~\ref{fgr:CGstr}(b-d), exemplifying higher densities, show that as $\rho\sigma_0^3$ increases, the $\mathbf{CP}$ and $\mathbf{CM}_p$ systems more and more deviate from the structure of $\mathbf{RC}$. The deviation is strongest to $\mathbf{CP}$ in which the structure is more pronounced, i.e. higher peaks and correlations, due to the missing softness in the pair potentials.  Only the $\mathbf{CM}_N$ model, which incorporates the density-dependent emergent distributions, can still describe the RC structure well up to moderate densities, $\rho\sigma_0^3=0.57$. Hence, we conclude that on the structural level the RC model gives rise to many body correlations which are very different than the conventional \cmp model, but can be approximately captured by a monodisperse model such as \cmn, when the emergent distributions are known. 

\subsection{Translational diffusion}
\noindent

\begin{figure}[htbp!]
 \centering
 \resizebox{!}{0.5\textwidth}{\input{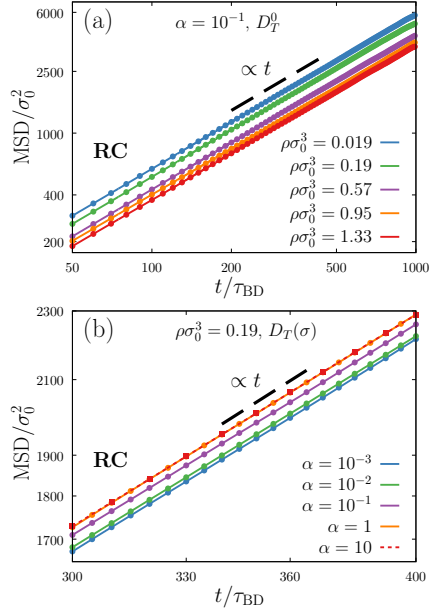}}
 \caption{{\footnotesize Representative positional mean-square displacements (MSDs) in the $\mathbf{RC}$ systems. They indicate diffusive behavior ($\propto t$, black dashed line). (a) With $\alpha=10^{-1}$ and constant friction ($D_T^0$) for varied $\rho\sigma_0^3$. (b) With $\rho\sigma_0^3=0.19$ and instantaneous size dependent friction ($D_T(\sigma)$) for varied $\alpha$.}}
 \label{fgr:MSD}
\end{figure}
\begin{figure}[htbp!]
 \centering
 \resizebox{!}{0.5\textwidth}{\input{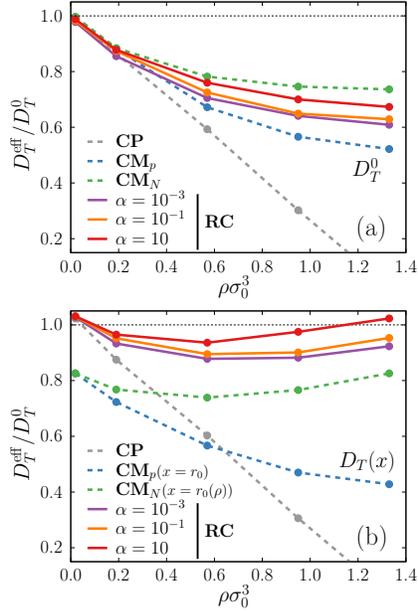}}
 \caption{{\footnotesize Dependence of the translational diffusion coefficient $D_T^{\mathrm{eff}}$ with number density $\rho\sigma_0^3$ for various systems, (a) with $D_T=D_T^0$ and (b) with $D_T=D_T(x)$ where $D_T(x)$ has the same meaning as in Tab.~\ref{tbl:DTrans} unless specified in plot legends. Note that with $D_T=D_T^0$, $D_T^{\mathrm{eff}}$ monotonically decreases with $\rho\sigma_0^3$ for all systems. For $D_T=D_T(x)$ in panel (b), however, $\mathbf{RC}$ systems show non-monotonic changes with $\rho\sigma_0^3$, only captured by the \cmn system.}}
 \label{fgr:Deff}
\end{figure}
\begin{figure}[htbp!]
 \centering
 \resizebox{!}{0.3\textwidth}{\input{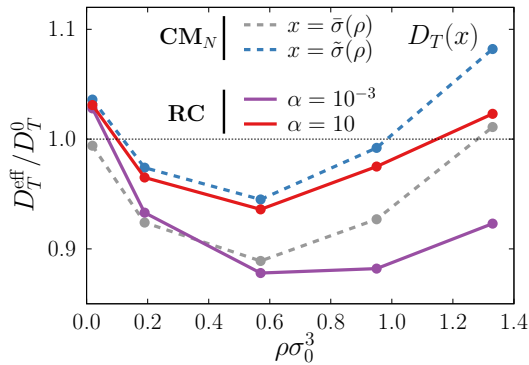}}
 \caption{{\footnotesize Translational diffusion coefficients $D_T^{\mathrm{eff}}$ for $\mathbf{RC}$ systems can be closely approximated at all number densities $\rho\sigma_0^3$ by $\mathbf{CM}_N$ systems by using single particle diffusion constants $D_T(\bar{\sigma}(\rho))$ (eq.~\ref{eqn:SigmaBar}) or $D_T(\tilde{\sigma}(\rho))$ (eq.~\ref{eqn:SigmaTilde}) in the BD equations of motion.}}
 \label{fgr:Deff2}
\end{figure}
%


The translational MSDs of the RCs are plotted in Fig.~\ref{fgr:MSD} and show perfect diffusive behavior in the long-time limit, for both constant single-particle diffusion constant $D_T^0$ (panel (a)) and size-dependent $D_T(\sigma)$ (panel B).  As we see already here, the RC diffusion constant is a function of $\alpha$, that is the time scale of dynamical coupling between size fluctuations and translational dffusion influences the systems' dynamics. The derived effective diffusion constants for the RC systems are summarized in Tab.~\ref{tbl:DTrans} for all densities and $\alpha$ values, as well as plotted in Fig.~\ref{fgr:Deff} versus density for selected $\alpha$. 

In Fig.~\ref{fgr:Deff}(a) we observe that the effective diffusion monotonically decreases for fixed $D_T^0$ with increasing densities for all systems due to crowding effects. There is a significant influence of the dynamical coupling $\alpha$ in the RC systems, but the effects are small, less than 10\%. A larger $\alpha$ (faster property dynamics) leads to faster translational diffusion. This implies that property fluctuations, beyond any Stokes size effect, can accelerate the diffusion in crowded environments. Size fluctuating particles can probably easier leave crowded environments and local cages by a spontaneous size fluctuation which leads to a smooth escape. Density effects on the emergent distributions probably also play a role, as indicated by the variation in size relaxation times discussed above and summarized in Tab.~II.  The conventional monodisperse systems \cmp and \cmn are close to the RC behavior, but cannot exactly match them. The particles in the \cmp model are slower than the RCs for all densities, while the \cmn is always faster, which must be attributed to the difference in their effective pair potentials, $v$ and $\tilde v$, respectively.  If we look at the {\bf CP} case, the situation changes dramatically at large densities: because no compression is possible, the effective diffusion decreases much more massively with density in the {\bf CP} system. 

In the more realistic scenario of variable $D_T(\sigma)$ according to Stokes friction, plotted in panel Fig.~\ref{fgr:Deff}(b), the effective diffusion of the RCs rises again for larger densities, that is, behaves non-monotonically. The reason is that the compression of the RCs leads to overall faster diffusion, apparently overcompensating the crowding effects.  The non-monotonicity in diffusion of the RCs can only be captured by the \cmn model which respects the density-dependent property distribution $N(\rho)$ and uses the effective Hertzian interaction radius $r_0(\rho)$ as the Stokes radius in $D_T(x=r_0)$. However, here an interesting and eye-catching effect appears: the diffusion in the low-density limit ($\rho\rightarrow 0$) in the \cmp and \cmn systems is systematically wrong by about 20\%. The reason is that $r_0$ is a result from coarse-graining the pair structure and only reflects the size in terms of the effective pair potential in the {\bf CM} systems. It is in that sense not related to the single particle size which still is $\sigma_0$ (in the LDL). Typically, we find that $r_0$ is about 20\% larger than $\sigma_0$. {\it Hence, this is a systematic error we often do in conventional coarse-graining because we estimate particle size based on the definition of the pair potential and not the single particle Stokes size.} (It is of course irrelevant if we are not interested in dynamical properties).  

This systematic error can be corrected by using a different definition for the effective Stokes radius $x$ in $D_T(x)$ for the {\bf CM} systems. We tested either using single particle diffusion constants $D_T(\bar{\sigma}(\rho))$  based on averaging the friction ($\propto \sigma$) according to eq.~(\ref{eqn:SigmaBar}) or $D_T(\tilde{\sigma}(\rho))$ averaging the diffusion ($\propto \sigma^{-1}$) according to eq.~(\ref{eqn:SigmaTilde}). We focus on the \cmn model where the averages are performed using the emergent distributions. The results are shown in Fig.~\ref{fgr:Deff2}: importantly, the \cmn model now describes the diffusive dynamics of the RC systems quite well, in particular for low to intermediate densities. Interesting is that in this density region, the two different definitions of the average Stokes radius match either the slower limit $\alpha = 10^{-3}$ or the faster limit $\alpha = 10$, respectively. In the faster limit, the properties relax quickly and particles appear on the diffusion time scale as average particles of one size. Here, the mean of the friction (according to the mean Stokes radius) seems appropriate to model the diffusive dynamics. For the slower limit, however, the property relaxation is slow on the diffusion time scale and there is a polydispersity in the particle's diffusion distribution. In this case, it seems more appropriate to work with the mean diffusion in the \cmn model. This should be inspiring for more theoretical studies to rationalize these observations more mathematically. 

\begin{table*}[!htb]
  \begin{center}
  \begin{tabular}{@{}|M{15mm}|M{13mm}|M{13mm}|M{13mm}|M{13mm}|M{13mm}|M{13mm}|M{13mm}|M{13mm}|M{13mm}|M{13mm}|M{13mm}|@{}}
    \hline \hline
    \multirow{3}{*}{$D_T$ type} & \multirow{3}{*}{$\rho\sigma_0^3$} & \multicolumn{5}{c|}{$\mathbf{RC}$ } & \multirow{3}{*}{$\mathbf{CP}$} & \multirow{3}{*}{$\mathbf{CM_P}$} & \multicolumn{3}{c|}{$\mathbf{CM}_N$} \\
    \cline{3-7} \cline{10-12}
     &  & \multicolumn{5}{c|}{$\alpha$} &  &  & \multicolumn{3}{c|}{$x$} \\
    \cline{3-7}  \cline{10-12}
     &  & $10^{-3}$ & $10^{-2}$ & $10^{-1}$ & $1$ & $10$ &  &  & $r_0(\rho)$ & $\bar{\sigma}(\rho)$ & $\tilde{\sigma}(\rho)$\\
    \hline \hline
    \multirow{5}{*}{$D_T^0$} & 0.019 & 0.978 & 0.979 & 0.982 & 0.989 & 0.988 & 0.985 & 0.996 & 0.996 &  & \\
    \cline{2-10}
     & 0.19 & 0.855 & 0.864 & 0.874 & 0.872 & 0.878 & 0.865 & 0.882 & 0.884  &  & \\
     \cline{2-10}
     & 0.57 & 0.705 & 0.714 & 0.725 & 0.758 & 0.760 & 0.593 & 0.672 & 0.782  &  & \\
     \cline{2-10}
     & 0.95 & 0.641 & 0.639 & 0.649 & 0.674 & 0.700 & 0.302 & 0.566 & 0.746  &  & \\
     \cline{2-10}
     & 1.33 & 0.609 & 0.617 & 0.629 & 0.662 & 0.673 & 0.026 & 0.522 & 0.736  &  & \\
    \hline \hline
    \multirow{5}{*}{$D_T(x)$} & 0.019 & 1.028 & 1.029 & 1.030 & 1.035 & 1.031 & 1.023 & 0.826 & 0.826 & 0.994 & 1.036 \\
    \cline{2-12}
     & 0.19 & 0.933 & 0.940 & 0.952 & 0.959 & 0.965 & 0.875 & 0.723 & 0.768  & 0.924 & 0.974 \\
     \cline{2-12}
     & 0.57 & 0.878 & 0.883 & 0.895 & 0.924 & 0.936 & 0.603 & 0.567 & 0.739  & 0.889 & 0.945 \\
     \cline{2-12}
     & 0.95 & 0.882 & 0.887 & 0.901 & 0.941 & 0.975 & 0.306 & 0.470 & 0.766  & 0.927 & 0.992 \\
     \cline{2-12}
     & 1.33 & 0.923 & 0.929 & 0.953 & 0.975 & 1.023 & 0.030 & 0.428 & 0.826  & 1.011 & 1.082 \\
    \hline \hline
  \end{tabular}
  \end{center}
  \caption{{\footnotesize Translational diffusion coefficients ($D_T^{\mathrm{eff}}$) calculated from MSDs. All values are in units of $D_T^0$. 
  The property diffusion is given by  $D_{\sigma} = \alpha D_T^0$.  Simulations are either conducted with constant diffusion $D_T^0$ or size-dependent diffusion $D_T(x)$. For $\mathbf{RC}$, $\mathbf{CP}$, $\mathbf{CM}_p$, and $\mathbf{CM}_N$ systems $x$ represents $\sigma$, $\sigma$, $r_0$, and $r_0(\rho)$ respectively. In the \cmn system we also employed the average values $\bar \sigma$ and $\tilde \sigma$ from distribution-averaging the sizes or inverse sizes, cf.~eqs.~(\ref{eqn:SigmaBar}) and (\ref{eqn:SigmaTilde}), respectively. }}
  \label{tbl:DTrans}
\end{table*}
\section{Conclusion}

In summary, we have studied the structure and dynamics of bulk suspensions of responsive colloids (RCs), featuring an internal degree of freedom (a property) by using overdamped Brownian Dynamics simulation. As a case study we specifically studied a model of soft hydrogels where the pair potential is a Hertzian potential and the internal degree of freedom is the colloid size fluctuating in a harmonic potential (the parent distribution). We compared the results to interesting reference systems, such a conventional polydisperse (CP) colloids (with frozen polydispersity, i.e., frozen initial properties) in the canonical ensemble and conventional monodisperse (CM) systems, for which the property distributions were integrated out. 

We found large differences in the liquid structure between the RC and CP models, essentially due to the absence of the responsive properties in the CP systems which make them significantly more packed and stronger correlated at elevated densities than the RC and CM systems. As expected, the CM model based on the conventional coarse-graining in the low density limit (CM$_p$, based on the parent distribution) performed well for low densities, while at moderate and high densities larger differences were observed in comparison to the RC systems. However, the structure in the latter could be approximately captured at higher densities by a phenomenologically improved CM model, CM$_N$, where the effective pair potential included the effects of the emergent property distributions. Importantly, we can thus conclude that on the structural level the RC model gives rise to many body correlations which are very different than the conventional CM$_p$ model, but can be approximately captured by a monodisperse model such as CM$_N$, when the emergent distributions are known. We took the latter from our own simulations but in principle one should be able to develop theories for it, at least on a perturbation level of theory. 

Regarding dynamics, we found that the time scale of property relaxations is density-dependent and has significant influence on the translational diffusion of the RCs (while it has no effect on equilibrium structure, as it should be). In particular, the RC diffusion was interestingly always close to the free diffusion of the particles due to colloid compression effects at high densites, counter-balancing crowding effects, in contrast to the CP or monodisperse CM$_p$ systems.  The mean translational diffusion of the RCs could be well reproduced using an effective single-particle diffusion constant in the CM$_N$ model in which the average, either of friction or diffusion, was based on the emergent distributions.  These observations imply that property fluctuations significantly accelerate/modify the translational diffusion in crowded environments. Recently, similar phenomena were reported for suspensions of highly deformable ring polymers in 2D, which can be also modelled as Hertzian Disks~\cite{ZaccNatPhys2019}. 
Importantly, the CM$_N$ model seems to capture these dynamical effects using effective (density-dependent) diffusion constants only. In future, more complex dynamics would be interesting to study by considering particles with non-linear elastic responses, or even polymodal internal distributions~\cite{Lin:PRE,briels} including internal switching~\cite{moncho}. 
\\
\begin{acknowledgments}
We thank Benjamin Rotenberg, Nils G\"oth, and Yi-Chen Lin for insightful discussions. This project has received funding from the European Research Council (ERC) under the European Union's Horizon 2020 research and innovation programme (grant agreement Nr. 646659). The authors acknowledge support by the state of Baden-W\"urttemberg through bwHPC and the German Research Foundation (DFG) through grant no INST 39/963-1 FUGG (bwForCluster NEMO). 
\end{acknowledgments}

%

\end{document}